\documentstyle[twoside, epsf]{article}

\input ibvs2.sty

\begin{document}
\IBVShead{5xxx}{00 Month 200x}

\IBVStitle{Spectroscopic binaries in the open cluster Trumpler 16 revisited}

\IBVSauth{LUNA, G.J.$^1$; LEVATO, H.$^2$; MALARODA, S.$^2$; GROSSO, M.$^2$}

\IBVSinst{Instituto de Astronomia e Geofisica, USP, Rua de Mat\~ao 1226, S\~ao Paulo, Brazil, e-mail: gjmluna@astro.iag.usp.br}
\IBVSinst{Complejo Astron\'omico El Leoncito, Av.Espa\~na 1512 Sur, San Juan, Argentina}
\IBVSkey{spectroscopy}
\vspace{1cm}

\IBVSabs{We have recalculated the orbits of 3 binary systems, members of the open cluster Trumpler 16 in the Carina region. We have added new observations to those available in the literature. Also, we have also tried to find the orbital periods for two suspected binary systems in the same cluster. We have improved considerably the precision of some of the orbital parameters.} 
\begintext

The open cluster Trumpler 16 is near the Carina nebulae (NGC 3372). There is a large amount of literature on the clusters in this region. The spectral morphology of members of Trumpler 16 has been studied by Levato \& Malaroda (1982) and references therein. The cluster distance has been
determined using different methods, and there is an agreement around 3.5
Kpc. This value is obviously dependeds on the interstellar absorption which may be abnormal for this cluster.
Trumpler 16 is a rich cluster with a large variety of objects like WR
stars, O3 stars, spectroscopic binaries and the singular $\eta$ Carinae. The
age estimated for Trumpler 16 is aproximately 10$^7$ years (Massey \& Johnson, 1993).

The study of spectroscopic binary systems is needed in order to tackle one of
the most classic problems in astronomy namely the stellar mass determination. A good knowledge of mass for individual stars allows its calibration with
other stellar parameters which are most easily observable, and also permits to test current stellar evolution theories.
 Spectroscopic binaries in open clusters are important in yet another context. It has been proposed that the average projected rotational velocity of the stars in an open cluster depends on its content of close spectroscopic binaries and magnetic chemically peculiar stars (Abt \& Sanders, 1973). Both tidal and magnetic braking are responsible for reducing the axial rotation. (Levato et al, 1987; Abt et al. 1973)

We used the Jorge Sahade-2.1m Telescope at CASLEO, San Juan, Argentina. A REOSC echelle spectrograph was employed during the following nights: February 27th 1996, from April 21st to 23rd 2000 and from March 10th to 11th 2001. The spectra centered on the blue wavelength region were recorded on a TEK1024 CCD and the one pixel resolution is 0.14 \AA.
The usual flat-fields and bias frames were obtained each night and the wavelength calibration was done using a Th-Ar lamp. The reduction was made with the standard procedure, using IRAF\footnote{IRAF is distributed by the National Optical Astronomy Observatories, operated by AURA, Inc., under cooperative agreement with the NSF. }

We have selected five stars for this project, probable members of Trumpler 16, from Levato et al. (1991). Three of them were chosen because the number of observations in their radial velocity curves was smaller than ten, and the other two because the authors considered them as radial velocity variables. 
To derive the radial velocities we have measured the Doppler shifts of H, He I, He II, Si IV, C IV, N III, C III lines. (Walborn et al. 2000), and applied an heliocentric correction to the measurements.
The observational results for the three stars for which we have recalculated the orbits can be seen the Table 1, where we have indicated in successive columns the Julian Date, the average radial velocity for each spectra, the number of lines measured, and the probable errors. The stars are identified by their HD number or Feinstein, Marraco, and Muzzio (1973) numbers. 

\vspace{0.2cm}
\centerline{Table 1: Radial Velocity Observations} 
\begin{center}
\small
\begin{tabular}{|c|c|c|c|c|}
\hline
\hline
\bf{Identification} & \bf{Julian Date} & \bf{RV} & \bf{n} & 
\bf{P.E.} \\ 
&  & \bf{(Km/s)} & \bf{n} & \bf{(Km/s)} \\ 
\hline
& 2450141.73 & -85.39 & 6 & 2.0 \\ 
\  & 2451656.52 & 18.46 & 7 & 6.2 \\ 
\ \bf{\#112}  & 2451657.52 & 54.45 & 8 & 5.8 \\ 
\  & 2451657.61 & 50.24 & 7 & 7.0 \\
   & 2451658.57 & -30.15 & 8 & 10.7 \\ 
\hline
\  & 2451979.76 & -76.7 & 9 & 11.5 \\
\  & 2451979.78 & -118.91 & 4 & 15.5 \\
 \bf{HD93161} & 2451979.79 & -91.11 & 12 & 8.6 \\
\ & 2451980.61 & -105.69 & 6 & 8.8 \\ 
\  & 2451980.65 & -96.64 & 8 & 12.3 \\
& 2451980.77 & -97.58 & 10 & 9.8 \\  
\hline
& 2451979.74 & -181.2 & 7 & 6.1 \\
 & 2451979.77 & -174.7 & 7 & 12.2 \\  
 \bf{\#104}  & 2451980.57 & -19.7 & 7 & 5.8 \\  
& 2451980.63 & 2.3 & 5 & 10.3 \\
\  & 2451980.74 & 109.2 & 7 & 18.6 \\
\  & 2451980.79 & 123.4 & 6 & 25.6 \\  
\hline
\hline
\end{tabular}
\end{center}
\vspace{0.5cm}

We have added the new observations presented in Table 1 to those published by Levato et al (1991) and searched for periods, using Morbey's code (Morbey, 1978) and, when successful, computed orbital elements starting from the results from Levato et al. (1991) improving them with the code of Bertiau \& Grobben  (1969). 

For HD93161, \#112 and \#114 stars, we found new orbital parameters, while for \#10 and \#110, we could not find any significant evidence of variability. We have applied an analysis of variance test (Conti et al. 1977) which confirms that the distribution of the observations for stars \#110 and \#10 do not depart significantly from a random one.
The new orbital elements for stars HD 93161, \#112, and \#114 are shown in Table 2.

\vspace{0.5cm}
\centerline {Table 2: Orbital Elements obtained in this work.} 
\begin{tabular}{|c|c|c|c|}
\hline
\hline
\bf{Element} & \bf{HD93161} & \bf{\#104} & \bf{\#112} \\ 
\hline
\bf{a \textit{sin i }(Km)} &  3.72 x 10$^{6}$ & 4.03 x 10$^{6}$ &  3.67 x 10$^{6}$  \\ 
\hline
\it{K (Km/s)} & 50.9$\pm $5.1 & 162.8 $\pm $ 11.4 & 86.0 $\pm $ 4 \\ 
\hline
\bf{e} &  0.309 $\pm $0.116 &  0.156 $\pm $0.057 &  0.249 $\pm $ 0.026 \\ 
\hline
$\mathbf{\omega ({{}^{\circ }})}$ & 184.4 $\pm $ 17.8 &  85.7 $\pm $ 28.4 &  286.6 $\pm $8.8 \\ 
\hline
\bf{T}$_{0}$(J.D.)(2.400.000+) &  45778.39 $\pm $ 0.22 &  45779.37 $\pm $ 0.13 & 45773.02  $\pm $ 0.1 \\ 
\hline
\bf{P}(days) & 5.60486 $\pm $ 9x10$^{-5}$ &  1.82303 $\pm $1x10$^{-5}$ & 4.07997 $\pm $2x10$^{-5}$ \\ 
\hline
\bf{V}$_{0}$ (Km/s) &  -44.5 $\pm $ 3.4 & -32.7  $\pm $ 7.5 &  -19.1  $\pm $ 1.7 \\ 
\hline
\bf{Mass Function} & 0.06 $\pm $ 0.03 &  0.78 $\pm $ 0.18 &  0.24 $\pm $ 0.02 \\ 
\hline
\hline
\end{tabular}
\vspace{0.5cm}

Figures 1, 2 and 3 show the radial velocity curves for \#104,
\#112 and HD93161 respectively.

\IBVSfig{10cm}{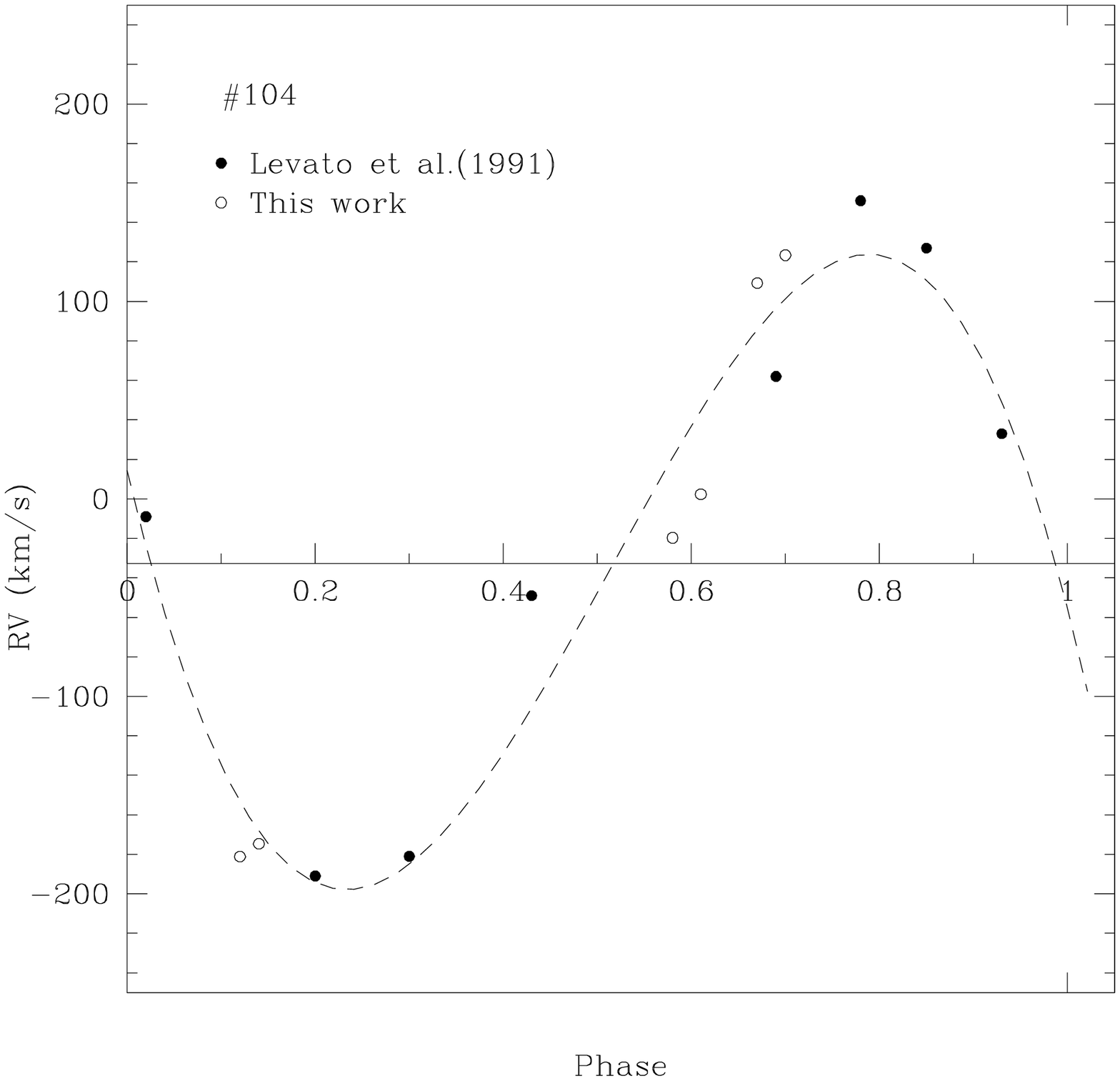}{Radial velocity curve of $\#$104}

\IBVSfig{10cm}{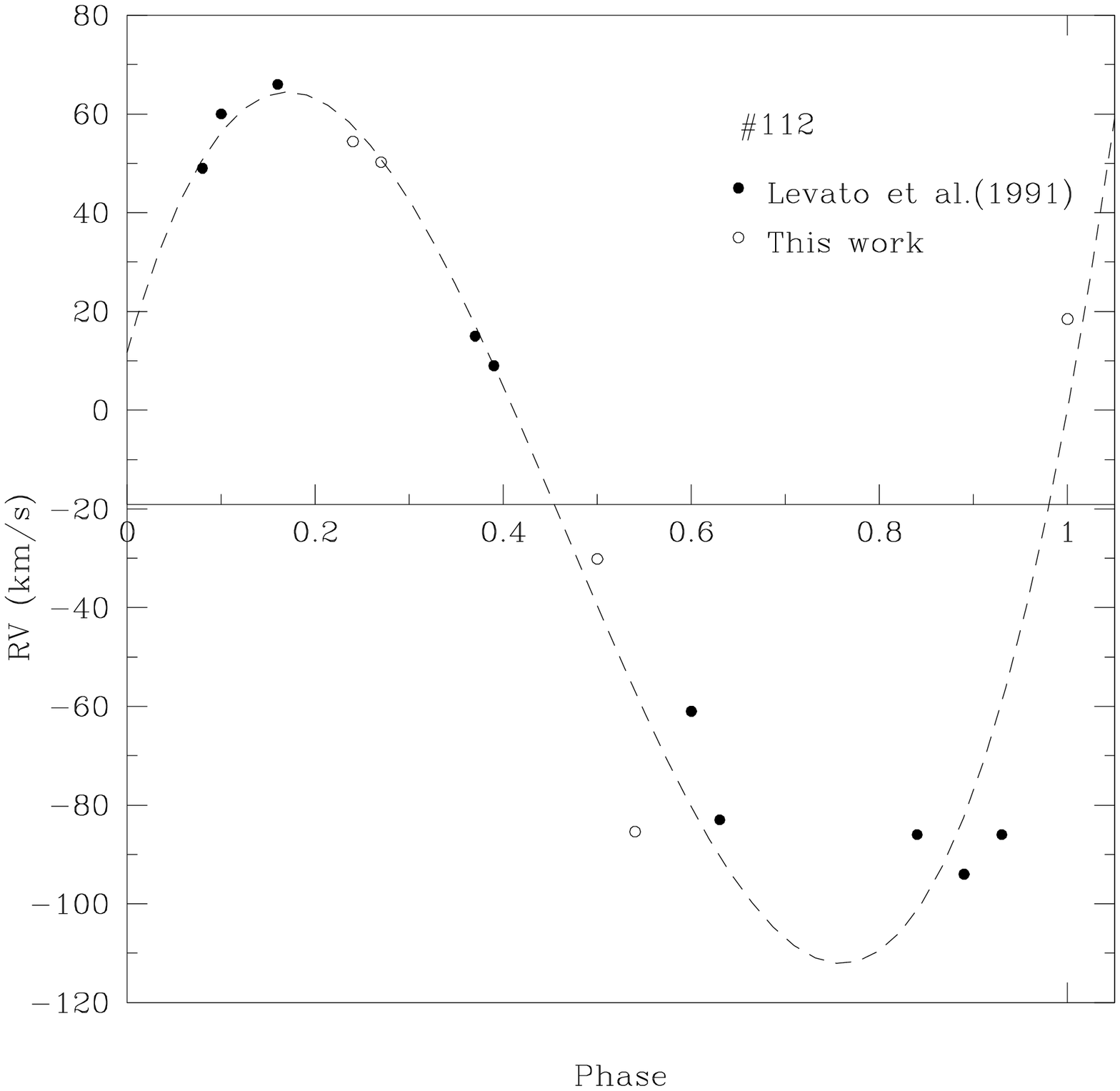}{
Radial velocity curve of $\#$112}

\IBVSfig{10cm}{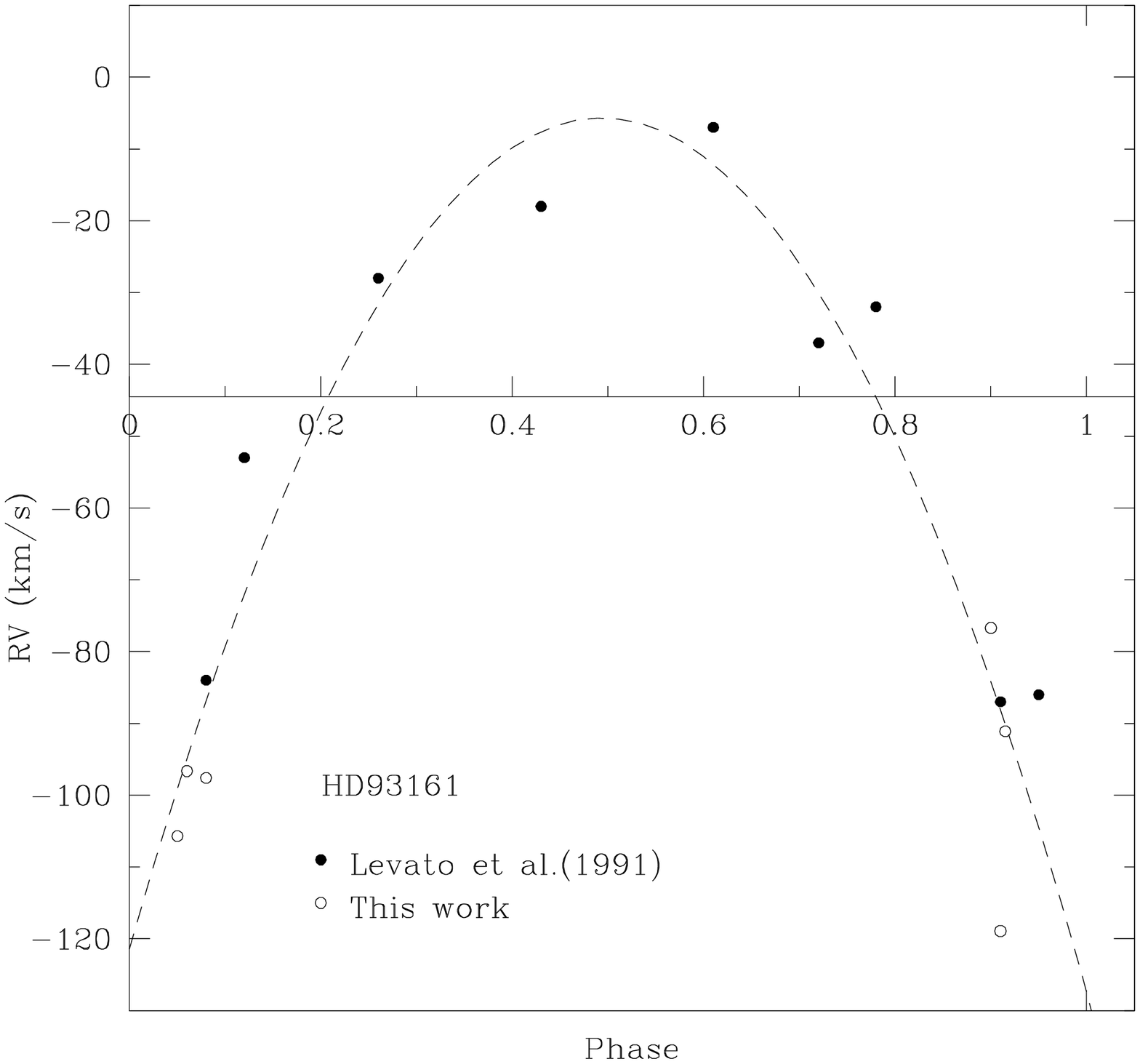}{Radial velocity curve of HD93161}

\vspace{1cm}

We have recalculated new orbital elements for three spectroscopic binary systems which belong to the open cluster Trumpler 16 and obtained the mass function for these three systems. Also we could not find significant radial velocity variations for \#110 and \#10 of the same cluster.

Support from Universidad Nacional de San Juan and CASLEO is deeply appreciated.

\references

Abt H.\& Sanders W., 1973, {\it ApJ}, {\bf 186}, 177 

Bertiau F. C. \& Grobben J., 1969, {\it Specola Astr. Vaticana Ric. Astr.}, {\bf 8}, No.1 

Conti P., Garmani C. \& Hutchings J., 1977, {\it ApJ}, {\bf 215}, 561

Feinstein A., Marraco H. G. \& Muzzio J. C., 1973, {\it A\&A}, { \bf 12}, 331 

Levato H., Malaroda S., 1982, {\it PASP}, {\bf 94}, 807 

Levato H., Malaroda S., Morrell N., Garcia B., Hern\'{a}ndez C., 1991,
{\it ApJS}, {\bf 75}, 869 

Levato H., Malaroda S., Morrell N. \& Solivella G., 1987, {\it ApJS}, {\bf 64}, 487

Massey, P. \& Johnson, J., 1993, {\it Astron. J.}, {\bf 105}, 980

Morbey, Ch., 1978, {\it Pub. Dom. Ap. Obs.}, {\bf 15}, 105

Walborn N. R., Lennon D., Heap S., Lindler D. Smith L., Evans J., Parker J.,
2000, \it  {STSI Preprint Series}, 1437

\label{lastpage}

\endreferences

\end{document}